\documentclass{ws-ijmpd}

\usepackage[super,compress]{cite}

\usepackage{graphicx}   
\usepackage{amsmath}
\usepackage{amssymb}
\usepackage{amsfonts}
\usepackage[utf8]{inputenc}
\usepackage[T1]{fontenc}
\usepackage{mathrsfs}

\mathchardef\arr="017E 
\renewcommand\vec[1]{\setbox0=\hbox{$#1$}\lower2ex\hbox to 0pt{\hbox to \wd0{\hss$\arr\;$\hss}\hss}\box0}

\usepackage{pifont}

\usepackage[linktocpage,breaklinks]{hyperref}

\usepackage{enumitem}

\usepackage[usenames]{color}
\definecolor{ao(english)}{rgb}{0.93, 0.53, 0.18}

\usepackage[normalem]{ulem}

\newcommand{\be}{\begin{equation}}
\newcommand{\ee}{\end{equation}}
\newcommand{\bii}{\begin{itemize}}
\newcommand{\eii}{\end{itemize}}

\usepackage{amsfonts}  
\usepackage{amsmath, amssymb, stmaryrd}     

\usepackage{graphicx}   
\usepackage{amsmath}    
\usepackage{amssymb}    

\usepackage{booktabs, multirow, array}
\usepackage[table]{xcolor} 

\usepackage{amsthm}
\theoremstyle{plain}
\newtheorem{thm}{Theorem}[section] 
\theoremstyle{definition}
\newtheorem{defn}[thm]{Definition} 

\usepackage{pdflscape}
\usepackage{afterpage}

\usepackage{array}
\newcolumntype{C}[1]{>{\centering\let\newline\\\arraybackslash\hspace{0pt}}p{#1}}

\begin{document}

\markboth{C. Llinares}
{Simulation techniques for modified gravity}

%
\catchline{}{}{}{}{}
%

\title{Simulation techniques for modified gravity}

\author{Claudio Llinares}

\address{Institute for Computational Cosmology, \\Department of Physics, Durham University, \\Durham DH1 3LE, UK \\
Institute of Cosmology and Gravitation, \\University of Portsmouth, Dennis Sciama Building, \\Portsmouth PO1 3FX, UK\\
Instituto de Astrof{\'i}sica e Ci{\^e}ncias do Espa{\c c}o, Universidade de Lisboa, \\
Faculdade de Ci{\^e}ncias, Campo Grande, PT1749-016 Lisboa, Portugal\\
cllinares@fc.ul.pt}

\maketitle

\begin{history}
\received{24 April 2018}
\revised{9 October 2018}
\accepted{29 October 2018}
\end{history}

\begin{abstract}
The standard paradigm of cosmology assumes General Relativity (GR) is a valid theory for gravity at scales in which it has not been properly tested.  Developing novel tests of GR and its alternatives is crucial if we want to give strength to the model or find departures from GR in the data.  Since alternatives to GR are usually defined through nonlinear equations, designing new tests for these theories implies a jump in complexity and thus, a need for refining the simulation techniques.  We summarize existing techniques for dealing with modified gravity (MG) in the context of cosmological simulations.  $N$-body codes for MG are usually based on standard gravity codes.  We describe the required extensions, classifying the models not according to their original motivation, but by the numerical challenges that must be faced by numericists.  MG models usually give rise to elliptic equations, for which multigrid techniques are well suited.  Thus, we devote a large fraction of this review to describing this particular technique.  Contrary to other reviews on multigrid methods, we focus on the specific techniques that are required to solve MG equations and describe useful tricks.  Finally, we describe extensions for going beyond the static approximation and dealing with baryons.
\end{abstract}


\ccode{PACS Numbers: 02.60.Cb, 02.60.Lj, 02.70.-c, 04.50.Kd, 98.65.-r, 98.80.-k}


\section{Introduction}

Tracking the evolution of the matter distribution in the late universe is a multi- scale problem in both space and time.  For instance, the evolution of the largest voids in the universe have associated scales of tens of Mpc and gigayears.  On the contrary, typical scales associated with merger events in high density regions are 
\newpage
\noindent
kpc and megayears or event shorter.  Thanks to this, studying the evolution of the universe as a whole is a highly non-trivial task.  This chapter is about how to calculate the evolution of matter over the largest possible dynamical range (i.e. maximizing the difference between the largest and smallest simulated scales).  While these are the most precise calculations that can be done, they are also the most expensive and require the implementation of complex numerical techniques.

According to the scales involved, there are four ways in which we can study the evolution of the distribution of matter in the universe:
\begin{itemize}[leftmargin=*]
\item \textit{Background cosmology:}  the matter distribution is assumed to be uniform.  In this case there is no gravitational collapse and the only dynamics is associated with the expansion rate.  The resulting equations are ordinary differential equations which can be solved analytically or with Runge--Kutta methods.
\item \textit{Linear cosmology:}  the matter distribution is assumed to have perturbations which are small when compared to the background density.  This means that it is possible to linearize the equations with respect to the density perturbations.  In the case of General Relativity (GR), this approximation gives rise to ordinary differential equations for the density perturbations.  By using this approach, it is possible to obtain high accuracy predictions for the redshift $z=0$ power spectrum for scales that correspond to frequencies $k \lesssim 0.1~h/$Mpc.
\item \textit{Nonlinear cosmology:}  when the matter perturbations become of order one, it is not possible to linearize the equations with respect to the density any more.  Under these conditions, it is necessary to keep all the terms in an expansion with respect to the density.  However, as cosmological evolution is not affected by the presence of black holes, nonlinear terms can still be neglected in the velocities and the metric perturbations.  The resulting equations are typically partial differential equations in space and time, and complex simulation techniques are required.
\item \textit{Relativistic or post-Newtonian cosmology:}   including relativistic species or super-horizon scales requires the addition of higher order terms in the Einstein's equations.  The solution of these equations can be obtained with specialized codes that take into account post-Newtonian corrections \cite{2016JCAP...07..053A,2017JCAP...11..004A}.
\end{itemize}
This paper will deal with the third-class of calculations in a modified gravity (MG) context.  With few exceptions, existing MG codes can simulate only scalar tensor theories, which means that they follow only one additional degree of freedom.  This requires solving the equations of motion for this field in addition to the standard $N$-body equations.  The paper is structured according to the class of equations that describe this field.  We will describe the methods step-by-step increasing the complexity of the equations and thus of the required solvers.  A summary of this classification, as well as the methods that are commonly used and existing implementations are given in Table \ref{table:classification_methods}.  An in-depth discussion of the definition of the models and, in particular, of the recurrent concept of screening mechanism can be found in the introductory paper of this special issue.
\newpage
Dark energy models share similarities with models that are intended to explain the dark matter component of our universe (a.k.a. MOND).  For this reason, there is also a lot to learn from the implementation of these particular models.  We summarize in Table \ref{table:classification_mond} existing methods and implementations for this family of theories.

The robustness of the methods and their implementation can be quantified by comparing results with analytic or linear solutions.  However, due to the lack of analytic solutions in the fully nonlinear regime, different implementations of the models are validated also by comparing codes that were written independently by different authors.  Such a comparison was presented for instance in Refs.~\refcite{2013MNRAS.436..348P} and \refcite{2014A&A...562A..78L}.  Furthermore, a methodical analysis of the accuracy of these codes 
was presented in Ref.~\refcite{2015MNRAS.454.4208W}.  Several quantities were compared: power spectrum at different redshifts, 
  \begin{table}[t]
    
    \tbl{Clasification of modified gravity models and their implementations.  In all the cases $L$ represents a differential operator and $f$ is a scalar function.}
        {\begin{tabular}{ p{0.30\textwidth} p{0.17\textwidth} p{0.30\textwidth} p{0.17\textwidth}}
            \toprule
            Type of equation  &  Model  & Method & References \\
            \toprule 
            \multirow{2}{*}{\begin{minipage}[c]{0.3\textwidth}
                $L(\phi, \dot{\phi}, \ddot{\phi}) = f(\rho,\phi)$ \\
                $L,f$ linear
            \end{minipage}} 
            & Dynamical DE & Modified expansion & ART\cite{2003ApJ...599...31K} \\ 
            \cmidrule{2-4}
            & Coupled DE & Modified expansion + $G_{\mathrm{eff}}$ & Refs.~\refcite{2010MNRAS.403.1684B,2004PhRvD..69l3516M} \\ 
            \cmidrule{2-4}
            & Non-local & Modified expansion + $G_{\mathrm{eff}}$ & RAMSES \cite{2014JCAP...09..031B} \\
            \cmidrule{2-4}
            & Vector DE & Modified expansion, power spectrum and growth history & Ref.~\refcite{2012MNRAS.424..699C} \\
            \midrule 
            \multirow{2}{*}{\begin{minipage}[c]{0.4\textwidth}
                $L(\phi, \partial_x\phi, \partial^2_x\phi) = f(\rho, \phi)$ \\
                $L$ linear in $\partial^2_x\phi$. 
            \end{minipage}} 
            & $f(R)$ & Multigrid (Jordan frame) &  Ref.~\refcite{2008PhRvD..78l3523O} \\
            & & & ECOSMOG \cite{2012JCAP...01..051L} \\
            & & & MG-GADGET \cite{2013MNRAS.436..348P} \\
            \cmidrule{3-4}
            & & Multigrid (Einstein frame) & Isis \cite{2014A&A...562A..78L} \\
            \cmidrule{2-4}
            & Generalized chameleon & Multigrid & ECOSMOG \cite{2012JCAP...10..002B} \\
            \cmidrule{2-4}
            & Symmetron & Multigrid & AMIGA \cite{2012ApJ...748...61D} \\
            &  & & ECOSMOG \cite{2012JCAP...10..002B} \\
            & & & Isis \cite{2014A&A...562A..78L}\\
            \cmidrule{2-4}
            & Dilaton & Multigrid & ECOSMOG \cite{2012JCAP...10..002B} \\
            \midrule
            \multirow{2}{*}{\begin{minipage}[c]{0.4\textwidth}
                $L(\phi, \partial_x\phi, \partial^2_x\phi) = f(\rho, \phi)$ \\
                $L$ non-linear in $\partial^2_x\phi$
            \end{minipage}} 
             & DGP & Multigrid with smoothing & DGPM \cite{2009PhRvD..80d3001S} \\
            \cmidrule{3-4}
            & & Fourier with smoothing & Refs.~\refcite{2009PhRvD..80f4023K}, \refcite{2008PhRvD..77b4048L} \\
            \cmidrule{3-4}
            & & 
                     {\begin{minipage}[c]{0.3\textwidth}
                         FFT-relaxation with \\operator splitting 
                     \end{minipage} }
                     & Ref.~\refcite{2009PhRvD..80j4005C} \\
            \cmidrule{3-4}
            & & 
                     {\begin{minipage}[c]{0.3\textwidth}
                         Multigrid with operator \\splitting 
                     \end{minipage} }
                     & ECOSMOG \cite{2013JCAP...05..023L} \\
            \cmidrule{2-4}
            & Cubic Galileon & 
                     {\begin{minipage}[c]{0.3\textwidth}
                         Multigrid with operator \\splitting 
                     \end{minipage} }
                     & ECOSMOG \cite{2013JCAP...10..027B} \\
            \cmidrule{2-4}
            & Quartic Galileon & 
                     {\begin{minipage}[c]{0.3\textwidth}
                         Multigrid with operator \\splitting 
                     \end{minipage} }
                     & ECOSMOG \cite{2013JCAP...11..012L} \\
            \midrule
            $L(\ddot{\phi}, \dot{\phi}, \partial_x\phi, \partial_x^2\phi) = f(\rho,\dot{\phi},\phi)$ & Symmetron & Leap-frog & Solve \cite{2013PhRvL.110p1101L}; Isis \cite{2014PhRvD..89h4023L} \\
            \cmidrule{2-4}
            & $f(R)$ & Implicit & ECOSMOG \cite{2015JCAP...02..034B} \\
            \cmidrule{2-4}
            & Disformal & Leap-frog & Isis \cite{2016A&A...585A..37H} \\
            \bottomrule
          \end{tabular}   \label{table:classification_methods}}
          
  \end{table}

\newpage
\noindent velocity power spectrum, mass function, density and velocity dispersion profile of halos and field distributions.   As the base codes used for each MG implementation were different, the comparison was made on the prediction of the deviations from GR rather that the absolute quantities.  It was found that all the MG codes that took part in the experiment agree in these predictions up to an accuracy of 1\%.  In particular, this accuracy was found for the power spectrum of density perturbations in a range of scales that goes from below $k=0.1$ to $k\sim 7~h$/Mpc, which is consistent with the scales that will be included in the next generation of galaxy surveys.

I close this introduction by giving a warning note.  Perceptive readers may notice the absence of diagrams in the text.  This is actually deliberate and intended to force the readers to do the diagrams by themselves.  Different people will make the diagrams in different ways.  This will increase the variety of interpretations of the methods described here, which in turn, will increase the chances of new ideas to be triggered by this modest contribution.

\section{The $N$-Body Equations for Cold Dark Matter Simulations}
\label{section_nbody_eqs}

A change of the gravitational theory constitutes a change of one of the fundamental pillars on which cosmological simulations are built.  Thus, before searching for simulation methods for MG, it is important to define in a precise way what cosmological simulations are and which equations are involved.  This is not a trivial task in the most general case.  So in order to keep the discussion focused on the changes that are required to simulate a alternative gravitational models, and not to get lost in technicalities of more advanced simulations, we will restrict the discussion to the simplest type of cosmological simulations.  These are simulations that deal with non-relativistic colisionless fluid (i.e. cold dark matter).  Some details of what is necessay to simulate other kinds of fluids will be given in the last section of this paper.

\begin{table}[t] 
  \tbl{Clasification of the MOND family of modified gravity models and their implementations.}
  {\begin{tabular}{ p{0.17\textwidth} p{0.30\textwidth} p{0.2\textwidth}}
    \toprule
    Model  & Method & References \\
    \toprule
    AQUAL & Algebraic & Ref.~\refcite{2002MNRAS.331..909N} \\
    & & AMIGA \cite{2004MNRAS.347.1055K} \\
    \cmidrule{2-3}
    & Multigrid & Ref.~\refcite{1999ApJ...519..590B} \\
    & &  Ref.~\refcite{2007A&A...464..517T} \\
    & & AMIGA \cite{2008MNRAS.391.1778L} \\
    & & Solve \cite{2011PhDT.......253L} \\
    \cmidrule{2-3}
    & Multigrid (on spherical grid) & NMODY \cite{2009MSAIS..13...89L, 2011ascl.soft02001L} \\
    \cmidrule{1-3}
    QUMOND & Poisson based & Solve \cite{2011PhDT.......253L} \\
    & & Ref.~\refcite{2012MNRAS.421.2598A} \\
    & & RAyMOND \cite{2015MNRAS.446.1060C} \\
    & & POR \cite{2015CaJPh..93..232L} \\
    \cmidrule{1-3}
    Bi-metric & Multigrid & Solve \cite{2011PhDT.......253L} \\
    \bottomrule
  \end{tabular}
   \label{table:classification_mond}}
\end{table}

\newpage
The simplest definition we could imagine is as follows:
\begin{defn}
\label{definition_0}
A cold dark matter cosmological simulation is the solution of Newtonian equations of motion in an expanding universe for a continuous density field of nonrelativistic collisionless matter.
\end{defn}
While this definition might be good enough when applied to standard gravity simulations, it is important to be more precise in the MG case.  This is because there is no guarantee that the approximations that were made to derive the Newtonian limit of GR are valid in MG.  We need to keep track of all these approximations, for which it is necessary to derive the equations starting from basic principles.  For instance, for some specific MG models, the gravitational slip (i.e. the ratio between time and spatial scalar perturbations of the metric) is not zero.  In these cases, new equations will have to be solved to take into account the new degrees of freedom that may arise. The form of these new equations can be only derived in a fully relativistic set up, (i.e. starting from the basic principles defined by the Einstein--Hilbert action and making necessary approximations).  Thus, we re-define a cosmological simulation as follows:
\begin{defn}
\label{definition_1}
A cold dark matter cosmological simulation is the solution of Einstein's equation (or their modified versions) up to first-order in the metric perturbations and velocities.  Furthermore, the energy--momentum tensor of the matter component is assumed to be isotropic and to represent nonrelativistic collisionless matter. \nolinebreak
\end{defn}
With the exception of the code presented in Ref.~\refcite{2017arXiv170904703L}, the integration of the equations is performed in a cubic region of space assuming periodic boundary conditions.  Given this definition, it is now easy to derive the equations that need to be solved in the cosmological codes.  We describe the derivation for the standard gravity case first and then describe the modifications that should be made in this derivation when dealing with extended models.

With few exceptions, these equations are solved with $N$-body methods, which rely on the fundamental concept of $N$-body particle.  These techniques are based on the idea of discretizing the continuous density fields and describing them with a large set of particles.  Thus, the relatively small set of equations that define the physical model (in our case Einstein's equations) is substituted by a much larger set of equations that describe the trajectory of these particles.  Since there is not a one-to-one correspondence between the $N$-body particles and the physical particles that are associated to the continuous fields, the interactions between the $N$-body particles must be smoothed out.  See Ref.~\refcite{1988csup.book.....H} for a detailed discussion on the concept of $N$-body particles and methods.  During the rest of this paper we will use the terms \textit{cosmological simulation} and \textit{$N$-body simulation} as synonyms.  However, the distinction between the concept of simulation and the method that we use to run them should always be kept in mind.

\subsection{General relativity}
\label{section_gr}

The aim is to solve Einstein's equation for a cosmological nonrelativistic collisionless fluid.  In $N$-body simulations, the density of such a fluid is described by a set of particles.  So we need to re-write Einstein's equation in a way that its solution can be described by particles.  We do this by taking its divergence
\be
\nabla^a T_{ab} = 0, 
\ee
which describes the conservation of energy.  The energy--momentum tensor associated with a set of $N$-body particles is then assumed to be the addition of Dirac's $\delta$ functions associated with each particle.  This gives a set of geodesics equations for each particle:
\be
\label{eq_geo}
\frac{d^2x_n^a}{d\tau^2} + \Gamma^a_{bc} \frac{dx_n^b}{d\tau}\frac{dx_n^c}{d\tau}= 0, 
\ee
where $\Gamma^a_{bc}$ are the Christofell symbols, the index $n$ runs over all the particles and $\tau$ is proper time (see for instance Exercise 1.9 in Ref.~\refcite{2004sgig.book.....C}).  In order to get the equations for the $N$-body codes, we need to fix the reference frame, which can be defined by fixing the metric.  The most general perturbation of the Friedmann--Lema{\^i}tre--Robertson--Walker metric in Newtonian gauge is given by
\be
\label{metric}
ds^2 = a^2(t) \left[-\left(1+2\Psi\right) dt^2 -2B_idx^i dt + \left(1-2\Phi\right)\delta_{ij} dx^i dx^j + h_{ij}dx^i dx^j\right], 
\ee
where $t$ is conformal time and $\Phi$, $\Psi$, $B_{i}$, and $h_{ij}$ are scalar, vector and tensor perturbations.  Different codes use a different independent variable for the integration. Common choices are supercomoving time \cite{1998MNRAS.297..467M}, expansion factor $a(t)$ or conformal time.  The supercomoving time has the advantage of making the hydro-dynamical equations independent of $a$, which allows the implementation of standard algorithms defined on Minkowski spacetime.  With the exception of the works presented in Refs.~\refcite{2016JCAP...07..053A} and \refcite{2017JCAP...11..004A}, vector and tensor perturbations are neglected.  Substituting this metric in Eq.~(\ref{eq_geo}) and changing the parameter of the geodesics from proper time to conformal time gives the following equation for the spatial coordinates of each particle:
\be
\ddot{x}^i_n + H(a)\dot{x}^i_n - \left( \frac{\partial\Phi}{\partial x^i} \right)_n = 0, 
\ee
where the dots represent partial derivatives with respect to conformal time and $H(a) = \dot{a}/a$ and second-order terms in the velocities were neglected.  The treatment of the damping term in this equation (second-term in the left-hand side) can be simplified with the following definition:
\be
p^i_n \equiv a \dot{x}^i_n, 
\ee
\newpage
\noindent which gives
\begin{align}
\label{geo_1}
\dot{x}^i_n & = \frac{p^i_n}{a(t)}, \\
\label{geo_2}
\dot{p}^i_n & = a(t) \left( \frac{\partial\Phi}{\partial x^i} \right)_n, 
\end{align}
which are the Hamilton equations that can be found in $N$-body codes.  Note the presence of expansion factors in these equations.  As the universe expands, objects become farther and farther away from each other, with a consequent weakening of the gravitational force between them.  These factors take into account this effect.  It may be surprising that the effect appears in the so-called comoving coordinates defined by Eq.~(\ref{metric}), in which the background position of the objects does not change with time.  Even if the objects do not move, their interdistance (which is a covariant quantity and thus, independent of the reference frame), still increases with time, which results in smaller and smaller forces as time passes.\footnote{It is important not to confuse the \textit{comoving coordinates} with the \textit{comoving distance}.  The second is an alternative definition of distance used by cosmologists which is defined as the distance with respect to the distance that is calculated using the background metric.  Under this alternative definition, galaxies do not depart from each other because of the expansion, however, the effect of the expansion is still present in the Poisson and Hamilton equations.}

In order to complete the set of equations that will be solved with the $N$-body code, we need to find an equation for the force that appears in Eq.~(\ref{geo_2}).  This will be given by the Einstein's equations themselves\footnote{Note that I define the Einstein's equation as containing a cosmological constant.  This may open up the discussion about this equation representing GR or the $\Lambda$CDM cosmological model within GR.  Different authors have different opinions about these definitions.  In order to fix the terminology and to avoid getting lost in philosophical discussions about what is GR and what is modified gravity, I define GR as in Eq.~(\ref{einstein_eq}).  Everything else will be referred to as modified gravity.}:
\be
\label{einstein_eq}
G_{ab} + g_{ab}\Lambda = T_{ab}.
\ee
The Definition \ref{definition_1} says that we need to linearize these equations with respect to the metric perturbations and velocities (but not with respect to the density, which is treated nonperturbatively).  \footnote{The linearization of the velocities mentioned here should not be confused with the linearization that is usually made when solving the equations in the context of linear cosmology.  Once the equations are determined, not restrictions exist in the amplitude of the velocities when solving for them with $N$-body codes.}  So we need to write down the perturbed Einstein's equations for the metric (\ref{metric}).  In the GR case, the $ij$ component of Einstein's equations says that the two scalar perturbations are equal.  This is also valid in MG in the Einstein frame, but it is not necessary valid in the Jordan frame.  However, as the conformal factor that relates both frames is very close to one, it is usually assumed that both perturbations are equal (i.e. the present generation of simulation codes do not include gravitational slip).

\newpage
Neglecting vector and tensor perturbations gives rise to equations that depend only on one degree of freedom.  This means that all the information provided by Einstein's equations can be condensed into one equation.  It is costumary to obtain this information from the 00 component of Einstein's equations.  By substituting the metric given by Eq.~(\ref{metric}) in Eq.~(\ref{einstein_eq}), neglecting second order terms in $\Phi$, and substracting the zeroth-order equation (i.e. Friedmann equation) we obtain a generalization of Poisson's equation in an expanding universe:
\be
\nabla^2\Phi = 4\pi G a^2(t) \delta\rho, 
\label{poisson}
\ee
where we have assumed the following energy--momentum tensor:
\be
T_{ab} = (\rho_0 + \delta\rho) u_a u_b + (P_0 + \delta P) (g_{ab} + u_a u_b), 
\ee
where $u^a$ is the 4-velocity of the fluid, $\rho_0$ and $P_0$ are background density and pressure and $\delta\rho$ and $\delta P$ are perturbations on these quantities (which are not assumed to be small).  Furthermore, appropriate gauge choices have to be made to remove time derivatives of the metric perturbation.  See Refs.~\refcite{2011PhRvD..83l3505C}, \refcite{2012PhRvD..85f3512G}, \refcite{2015PhRvD..92l3517F} and \refcite{2017JCAP...12..022F} for a discussion on this issue.  The function $a(t)$ in Eqs.~(\ref{geo_1}), (\ref{geo_2}) and (\ref{poisson}) is obtained as a solution of the zeroth-order Einstein's equations (i.e. the Friedman equations):
\be
\frac{\dot{a}(t)}{a(t)} = H_0^2 \left[ \frac{\Omega_m}{a(t)^3} + \Omega_{\Lambda} \right].
\label{friedmann}
\ee
The routines that solve Eq.~(\ref{poisson}) are some of the more complex routines in $N$-body codes and will be the main focus of the sections that follow.  Several algorithms exist at present.  For instance, particle mesh (PM) codes solve explicitly Poisson's equation on a grid (through multigrid or Fourier-based methods) and estimate the force with discretization formulae for the derivatives.  The forces are then interpolated back to the particle positions using specific smoothing kernels.  The density that is required to solve the Poisson's equation is calculated from the particle distribution using the same type of interpolation.\footnote{We remind the reader that in order to ensure momentum conservation, the same kernel must be used to estimate the density on the grid and to interpolate the force to the particles' position.  The definitions of commonly used interpolation kernels can be found in Ref.~\refcite{1988csup.book.....H}.}

Alternatively, in the case of direct summation codes, Poisson's equation is integrated once analytically and thus, the gravitational force can be obtained by adding up individual contributions from each particle:
\be
\label{summation}
(\nabla\Phi)_n = 4\pi G \sum_m \sum_n \frac{M_n M_m}{|x^i_n-x^i_m|^2} \textbf{x}_{mn}, 
\ee
where the indexes $m$ and $n$ run over all the particles and $\textbf{x}_{mn}$ is a unitary vector between the particles $m$ and $n$.  In practice, these very expensive summations can be substituted by approximated summations in which the geometry of the long range interactions is simplified.  Furthermore, it is possible to combine direct summation and PM algorithms to construct even faster codes.  Finally, there are the so-called multipole methods, which rely on a multipolar expansion of Poisson's equation.

Note that direct summation and multipole algorithms can be implemented only because Poisson's equation is linear (i.e.~superposition principle holds).  In general, this property does not apply to equations that arise in MG models and thus, direct summation, multipole or Fourier techniques are not used for simulating MG.  The exact implementation of these methods in the standard gravity case is beyond the scope of this paper and can be found for instance in Refs.~\refcite{1998ARA&A..36..599B}, \refcite{2005CSci...88.1088B}, \refcite{2008SSRv..134..229D} or \refcite{2011ASL.....4..204B}.\footnote{Further information on how simulations work can be found in the presentation of individual codes listed in Tables \ref{table:classification_methods} and \ref{table:classification_mond} and references there in.  The ultimate way of understanding how an $N$-body code works is to write one from scratch.  This is not such as big task as it seems.  State-of-the-art codes typically contain more that 100.000 lines and require several years of development.  However, minimalistic implementations for training purposes, can consist only in 2000 to 3000 lines which can be easily written.}

To summarize, in the standard gravity case, it is possible to translate the definition of $N$-body simulation given at the beginning of this section into the following more practical definition.
\begin{defn}
  \textit{A standard gravity cold dark matter cosmological simulation is the solution of Eqs.~(\ref{geo_1}) and (\ref{geo_2}) for a large set of particles.  The force that appears in these equations can be obtained by either solving Eq.~(\ref{poisson}) on a grid (in the case of particle-mesh codes) or using Eq.~(\ref{summation}) (in the case of tree codes) together with Eq.~(\ref{friedmann}).}
\label{definition_2}
\end{defn}
The next section reformulates this definition for the MG models that were already implemented in codes and for which already exist nonlinear simulations.

\subsection{Modified gravity}
\label{section_mg}

Following the Definitions \ref{definition_1} and \ref{definition_2}, the implementation of a MG model in an $N$-body code has two distinctive stages.  First, the $N$-body equations have to be determined.  This means, that it is necessary to find a generalization of Eqs.~(\ref{geo_1}), (\ref{geo_2}), (\ref{poisson}) and (\ref{friedmann}).  The choice of these generalized equations is not unique.  For instance, very complex theories can be mapped into simple ones.  Furthermore, in several cases it is possible to approximate complex equations with simple ones for which simpler methods can be applied without loss of accuracy.  The decisions taken at this stage can make the difference between being able to simulate a model with minimal effort and not simulate it at all.  Once the equations have been fixed, it is possible to implement them in $N$-body codes.  In many cases the resulting equations largely differ from the type of equations that can be found in GR.  However, these equations may exist in different fields of physics.  Multidisciplinary collaborations and communication between different communities are crucial when it comes to the implementation stage.

\subsubsection{Scalar field theories}
\label{section:scalar_tensor}

Almost all the theories that have been simulated can be defined by adding a scalar field to the Einstein--Hilbert action.  There is almost infinite freedom in the way this can be done.  However, all this diversity can be condensed into four different properties of the action.  So the most general action that has been simulated is
\begin{align}
\nonumber
S = & \int \sqrt{-g} R d^4x + 
\int \sqrt{-g} \left[f(\partial \phi, \partial^2\phi) + V(\phi) \right] d^4x + \\
& \int \sqrt{-\tilde{g}} \mathcal{L}_M(m(\phi), \tilde{g}_{ab}(g_{ab}, \phi)) d^4x  = S_{G} + S_{KG} + S_{M} , 
\label{general_lagrangian}
\end{align}
where $S_{G}$ is the action that describes the geometry, $S_{KG}$ the one that gives the Klein--Gordon equation for an uncoupled field and $S_{M}$ is the matter action.  The four characteristics that define different classes of theories are as follows:
\begin{itemize}
\item $f(\partial \phi, \partial^2\phi)$: a free function of first and second covariant derivatives of the field $\phi$ (i.e. the kinetic term), which includes contractions between derivatives.
\item $V(\phi)$: a potential in which scalar field oscillates or roles.
\item $m(\phi)$: a coupling between the scalar field and the matter fields.
\item $\tilde{g}_{ab}(g_{ab}, \phi)$: a relation between the Jordan and Einstein frame metrics.
\end{itemize}
Different choices of these four ingredients give different families of theories.  The classification that is relevant for numericists is the one that highlights the numerical complexity of the resulting equations.  In this sense, the exact form of the free functions does not matter, but only its functional dependence.  The complexity varies from very simple models in which the scalar field is uncoupled and only affects the background evolution of the Universe up to theories that introduce nonlinear couplings with the matter fields which can lead to fully nonlinear partial differential equations in space and time. Table \ref{table:classification_scalar} summarizes this classification with increasing complexity of the equations and methods.  Note that this classification is inspired in both the original definition of the models as well as the numerical techniques that

\begin{table}[b] 
  \tbl{Classification of scalar field models that has been simulated.}
      {\begin{tabular}{ p{0.05\textwidth} C{0.21\textwidth} C{0.1\textwidth} C{0.15\textwidth} C{0.22\textwidth} p{0.14\textwidth}}
          \midrule \strut
          Class & Kinetic term & Potential & Coupling & Jordan metric & Model  \\
          \midrule
          1 & $(\partial\phi)^2$ & $V(\phi)$ & 1 & $g_{ab}$ & Ratta\&Peebels SUGRA     \strut \\
          \midrule
          2 & $(\partial\phi)^2$ & $V(\phi)$ & $m(\phi)$ (nonuniversal) & $g_{ab}$ & Coupled DE \\
          \midrule
          3 & $(\partial\phi)^2$ & $V(\phi)$ & 1 & $A(\phi)g_{ab}$ & Chameleon Symmetron Dilaton  \\
          \midrule
          4 & $(\partial\phi)^2$ & $V(\phi)$ & 1 & $A(\phi)g_{ab} + B(\phi)\phi_{,a}\phi_{,b}$ & Disformal \\
          \midrule
          5 & $(\partial\phi)^2 + \mathrm{higher~order}$ & $V(\phi)$ & 1 & $g_{ab}$ & Galileon \\
          \bottomrule
        \end{tabular}
        \label{table:classification_scalar}}
\end{table}

\newpage

\noindent have been applied.  However, the classification has some flexibility: theories from a given class may be mapped into theories that belong to different classes (see for instance Ref.~\refcite{2013PhRvD..87h3010Z} for a detailed discussion on the relation between classes 4 and 5 in Table \ref{table:classification_scalar}).  Implementation techniques that are applied to each class follow.

\smallskip

\noindent \textbf{Class 1. Unperturbed fields:}  The simplest imaginable scalar fields are the ones that have a standard kinetic term (i.e. $f=(\partial \phi)^2$) and are uncoupled.  As the field is not affected by metric nor matter perturbations, it is expected that it will not have spatial perturbations and thus, it will only evolve in the background.  The only effects that this will produce in the matter distribution are through the energy of the field, which will affect the expansion of the Universe.  Two modifications have to be made to the codes to include this class of models:  background expansion tables (which are needed because of the dependence of Eqs.~(\ref{geo_1}), (\ref{geo_2}) and (\ref{poisson}) on $a$) and the power spectrum that is required for generating the initial conditions (i.e. the growth factor).  Implementation of these models has no overhead with respect to GR simulations and thus it is possible to simulate these models with the same resolution that can be reached in state-of-the-art standard gravity simulations.

\smallskip

\noindent \textbf{Class 2. Fields with explicit coupling to dark matter:}  The simplest way to add a coupling to matter is to couple only the dark matter component and leave the baryons uncoupled.  This will allow the model to pass solar system tests and at the same time allow the use of very simple equations which do not include screening mechanisms.  The consequence that such coupling has in the $N$-body equations is the addition of an effective gravitational constant which is a function of the field plus an additional friction term.

\smallskip

\noindent \textbf{Class 3.  Fields with conformal coupling:} Giving a field dependence to the Jordan frame metric provides an extra term in the conservation of energy which acts as a fifth force in the geodesics equation (Eqs.~(\ref{geo_1}) and (\ref{geo_2})).  In addition to this, the equation of motion for $\phi$ becomes nonlinear.  Different choices of $V(\phi)$ and $A(\phi)$ give rise to different nonlinear terms which in turn give rise to different screening mechanisms.  Thanks to these screening mechanisms, these fields can be coupled to the baryons and at the same time pass solar system tests.  The standard kinetic term gives rise to a hyperbolic Klein--Gordon equation which can be solved with the methods described in Sec.~\ref{section:non_static}.  However, it is customary to assume the quasi-static approximation (i.e. to neglect the time derivatives of the field).  This will give rise to elliptic equations and enable us to implement the multigrid methods described in Sec.~\ref{section:multigrid}.

\smallskip

\noindent \textbf{Class 4. Fields with disformal coupling:}  The addition of derivatives of the field in the definition of the Jordan frame metric introduces a coupling with the time derivatives of the field in both the Klein--Gordon equation as well as in the geodesics.  This means that the equations contain terms such as $\dot{\phi}^2\rho$ and $\dot{\phi}^2\nabla\phi$.  These terms can only be treated with the nonstatic solvers described in Sec.~\ref{section:non_static}.

\newpage

\noindent \textbf{Class 5. Fields with non-standard kinetic terms:}  The addition of higher-order terms in the kinetic part of the scalar field action can give rise to fully nonlinear equations of motion for the field (i.e. the equations are nonlinear in the second-derivatives of the field).  These equations were included in the codes in the quasi-static limit using the multigrid techniques presented in Sec.~\ref{section:multigrid}.  The full nonlinearity makes the additional techniques presented in Sec.~\ref{section:fully_non_linear} mandatory.  Note that these models do not include coupling to matter (both explicitly or through the definition of the Jordan metric).  This means that the geodesics equations are unchanged and the effects of the scalar field on the matter distribution are through the energy of the scalar field which will appear as an additional term in the right-hand side of the Poisson's equation (Eq.~(\ref{poisson})).

Theories such as $f(R)$ are not included in Table \ref{table:classification_scalar} because, even if it is possible to map them into scalar-tensor theories, their definition have different motivation (i.e. they are not defined in the Einstein frame by adding a scalar field, but by adding a function of the Ricci scalar to the Einstein--Hilbert action in the Jordan frame).  Details on these other theories, are given below in independent sections.

\subsubsection{Nonlocal gravity}
\label{section:non_local}

A particular representation of the model can be defined by adding a term $m^2 R \Box^{-2}R$ to the Einstein--Hilbert action\cite{2014FoPh...44..213W, 2014PhRvD..90b3005M, 2014JCAP...06..033D} and was already implemented in the Ramses code\cite{2014JCAP...09..031B}.  It can be shown that this nontrivial nonlocal action, can be recast into a local action which includes two scalar fields.  The resulting equations of motion for these fields are linear and thus, it is possible to use the simple methods that are applicable to class 2 described in Sec.~\ref{section:scalar_tensor}.

\subsubsection{The $f(R)$ family of theories}
\label{section:fofr}

The action that defines this family of theories is usually given in the Jordan frame:  
\be
S_{f(R)} = \int \frac{\sqrt{-\tilde{g}}}{16 \pi G} \left[ R + f(R) \right] d^4x + S_m(\tilde{g}_{ab}, \psi),
\label{action_fr}
\ee
where $f$ is a free function and $\psi$ are matter fields.  The only model that has been simulated to date was proposed in Ref.~\refcite{2007PhRvD..76f4004H} and it is defined with the following function
\be
f(R) = -m^2\frac{c_1\left(\frac{R}{m^2}\right)^n}{c_2\left(\frac{R}{m^2}\right)^n + 1}.
\ee
The linearized Einstein's equation includes two scalar degrees of freedom, which means that in addition to a modified Poisson's equation, the model requires an extra equation to track this extra field.  These two degrees of freedom can be parametrized 
\newpage
\noindent as the usual gravitational potential $\Phi$ (i.e. the scalar perturbation of the metric defined by Eq.~(\ref{metric})) and the derivative of the free function itself $f_R \equiv \frac{df}{dR}$.  After minimizing the action (\ref{action_fr}) and assuming the quasi-static limit, we obtain the following linearized equations for these two degrees of freedom:
\begin{align}
\label{fr_1}
\nabla^2 \Phi & = \frac{16\pi G}{3}\delta\rho - \frac{1}{6} \delta R(f_R) \\
\label{fr_2}
\nabla^2 f_R & = -\frac{8\pi G}{3 c^2}\delta \rho  + \frac{1}{3 c^2} \delta R(f_R), 
\end{align}
where $\delta R(f_R)$ is the perturbation of the Ricci scalar with respect to the background cosmological value.  Its functional dependence with the field $f_R$ is nonlinear and given by
\be
\delta R(f_R) = \bar{R}(a) \left( \sqrt{\frac{\bar{f}_R(a)}{f_R}} - 1 \right), 
\ee
where $\bar{R}(a)$ and $\bar{f}_R(a)$ are background quantities, the free parameters of the model were fixed such that the expansion is consistent with observations (i.e. such that it gives an effective cosmological constant) and $n=1$. Under these conditions, the model has only one free parameter $f_{R0}$, which sets the normalization of the background field $\bar{f}_R(a)$.  The set of equations that are necessary to describe nonlinear cosmology is completed with the geodesics equations.  As these theories do not include additional fields besides the metric, the geodesics are unchanged and thus, the trajectory of the particles is determined by $\nabla \Phi$.  By combining Eqs.~(\ref{fr_1}) and (\ref{fr_2}) it is possible to see that the MG contribution to the acceleration is given by $c^2\nabla f_R/2$.

Equations (\ref{fr_1}) and (\ref{fr_2}) substitute Eq.~(\ref{poisson}) in the Definition \ref{definition_2}.  These two equations must be solved together, however, the second equation is independent of the metric perturbation $\Phi$ and thus, can be solved independently to obtain $f_R$ in every time step of the simulation.  After this is done, the field $f_R$ can be introduced in Eq.~(\ref{fr_1}) to obtain the metric perturbation and thus, the force that is required to move the particles.

A different implementation can be constructed by taking into account that the theory can be mapped into a theory that includes a chameleon field\cite{2008PhRvD..78j4021B}:
\be
\label{action_fr_einstein}
S_{f(R),E} = \int \frac{\sqrt{-g}}{16 \pi G} \left[ R + (\delta\phi)^2 +V(\phi) \right] d^4x + S_m(\tilde{g}_{ab}, \psi),
\ee
where $g_{ab}$ and $\tilde{g}_{ab}$ are Einstein and Jordan frame metrics respectively.  This mapping can be done by defining a scalar field $\phi$ such that 
\be
\exp\left(-\frac{2\beta\phi}{M_{\mathrm{P}}}\right) = f'(R) \equiv f_R, 
\ee
\newpage
\noindent where $M_{\mathrm{P}}$ is the Planck mass and the coupling constant is $\beta=1/\sqrt{6}$.  The relation between the Einstein and Jordan frame metrics is
\be
\tilde{g}_{ab} = \exp\left(-\frac{2\beta\phi}{M_{PL}}\right) g_{ab}.
\ee
In the new frame (i.e. the Einstein frame where $f(R)=0$), the equations for the two degrees of freedom (Eqs.~(\ref{fr_1}) and (\ref{fr_2})) are substituted with the following:
\begin{align}
\label{fr_eintein1}
\nabla^2 \Phi & = 4 \pi G\delta\rho - \rho_{f_R} \\
\label{fr_einstein_2}
\nabla^2 f_R & = -\frac{1}{a}\Omega_m H_0^2\left(\eta - 1\right)  + a^2\Omega_m H_0^2 \left[\left(1+4\frac{\Omega_\Lambda}{\Omega_m}\right)\left(\frac{f_{R0}}{f_R}\right)^{\frac{1}{n+1}}  - \left(a^{-3} + 4\frac{\Omega_\Lambda}{\Omega_m}\right)\right], 
\end{align}
where $\rho_{f_R} = \dot{f}_{R}^2 + |\nabla f_{R}|^2 + V_{\mathrm{eff}}(f_R)$ is the energy of the scalar field (which is assumed to be small) and $\eta$ is the matter density in terms of the background density.

As the part of the action (\ref{action_fr_einstein}) that defines the geometry is the same as in GR, the MG effects on the mater distribution appear as a fifth force in the geodesics equations:
\be
\frac{d^2{\bf x}}{d\tau^2} + \nabla\tilde{\Phi} + \frac{1}{2}\nabla f_R = 0, 
\ee
where $\tau$ and $\tilde{\Phi}$ are super-comoving time and potential, which are the variables that are used in the only code that includes this representation of the model \cite{2014A&A...562A..78L}.

Once a frame is chosen, we need to find a way of solving the field equations.  In both cases, the field $f_R$ satisfies the equation of a chameleon field.  This is a quasi-linear equation, which can be solved with the nonlinear multigrid methods described in Sec.~\ref{section:multigrid}.  In order to ensure that the field is positive and to increase the stability of the solvers, it is customary to solve the equations for a transformed field $u$, which is defined by
\be
f_R = - \frac{A}{a(t)^b} e^u, 
\label{change_fofr}
\ee
where the normalization $A$ and the dependence $b$ with the expansion factor was chosen differently by different authors.  While this change of variables provides robustness to the solvers, it forces them to evaluate an exponential function in every point of the grid and every iteration step.  This is a very expensive function to calculate and thus the implementation of this change reduces the performance of the solvers.  A workaround to this problem, which is based on a different change of variables that enables the implementation of an explicit solver, is described in Ref.~\refcite{2017JCAP...02..050B}.

The two formalisms described above are mathematically equivalent.  However, the numerical errors associated with the solution of these two set of equations may be different and thus, there is no guarantee that the end result of simulations ran in the Jordan or Einstein frame will be the same.  This has been tested in Ref.~\refcite{2015MNRAS.454.4208W}, were three different implementations of the model were compared (two of them written in the Jordan frame and one in the Einstein frame).  The differences between the predictions provided by these codes are below 1\%, which is enough for the accuracy required by present and near future surveys.

\subsubsection{Models with higher number of dimensions (DGP)}

Models defined in higher dimensional space-times, such as the DGP model, \cite{2000PhLB..485..208D} can be mapped into the scalar tensor theories described in Sec.~\ref{section:scalar_tensor}, which include the Vainshtein screening mechanism (class 5 in Table \ref{table:classification_scalar}) \cite{2009PhRvD..80d3001S,2009PhRvD..80l3003S,2013JCAP...05..023L}.  Once this is done, it is possible to apply the same techniques applied to fully nonlinear equations (see Sec.~\ref{section:multigrid} and in particular Sec.~\ref{section:fully_non_linear}).

\subsubsection{Vector fields}

The only vector-tensor theory that has been simulated is defined by the following action \cite{2012MNRAS.424..699C}:
\be
S = \int \sqrt{-g} \left[ - \frac{R}{16 \pi G} -\frac{1}{4} F_{ab}F^{ab} - \frac{1}{2}\left(\nabla_a A^a \right)^2 + R_{ab}A^a A^b \right]  d^4 x, 
\ee
where\footnote{Note that a different sign convention was used by these authors.}
\be
F_{ab} = \partial_a A_b - \partial_bA_a.
\ee
The equations of motion for the four extra degrees of freedom are
\be
\label{eq_vector}
\square A_a + R_{ab}A^b = 0.
\ee
As the part of the action that defines the geometry is the same as in GR, Einstein's equations are unchanged, with the exception of the addition of the energy--momentum tensor of the vector field in the source.  This extra term does not add numerical complexity.  However, the equation of motion for the vector (Eq. (\ref{eq_vector})) may be extremely difficult to solve in the most general case.

In the only existing simulations\cite{2012MNRAS.424..699C}, the perturbations of the vector field were not tracked and only the effects on the background expansion and the linear cosmology were taken into account.  This means that the equations Eqs.~(\ref{geo_1}), (\ref{geo_2}) and (\ref{poisson}) were kept as described in Definition \ref{definition_2}.  Only the solution of Eq.~(\ref{friedmann}) was changed while running the simulation.  In addition to this, the initial conditions were modified to take into account the presence of the vector field at high redshift.  These modifications apply to both the initial power spectrum as well as the growth rate.  In other words, these authors made appropriate approximations which map the theory into the class 1 defined in Table \ref{table:classification_scalar} of Sec.~\ref{section:scalar_tensor}.

\subsubsection{The MOND case}

From the point of view of late time cosmology, it is possible to classify MG theories into two large families which are motivated as alternative solutions to the two main open problems in cosmology:  dark matter and dark energy.  While this special issue focuses in the second class of theories, the equations associated with both families of theories share similarities.  Thus, there is a lot to learn from numerical implementations of solutions of the MOND equations, which are representative of alternatives to dark matter.

While there are several relativistic extensions, the original MOND idea is defined in a non-relativistic set up \cite{1983ApJ...270..365M}.  The theory substitutes the usual Newtonian potential with a modified potential, which is solution of 
\be
\nabla\cdot\left[ \mu\left(\frac{|\nabla\Phi_M|}{a_0}\right) \nabla\Phi_M \right]  = 4 \pi G \delta\rho, 
\label{mond_equation}
\ee
where $\mu$ is a free function and $a_0$ a free parameter which can be fixed, for instance, by requiring the theory to explain rotation curves of galaxies in the absence of dark matter.  A classical action for this equation can be found in Ref.~\refcite{1984ApJ...286....7B}.  The equation is quasi-linear and thus can be solved with the same multigrid methods that will be described in Sec.~\ref{section:multigrid}.

An alternative method commonly used to solve the MOND equation consists in equating the left-hand side of Eq.~(\ref{mond_equation}) with the left-hand side of the standard Poisson's equation.  Integrating once gives
\be
\mu\left(\frac{|\nabla\Phi_M|}{a_0}\right) \nabla\Phi_M + \nabla\times\textbf{k} = \nabla\Phi, 
\label{spherical_mond}
\ee
where the curl term $\textbf{k}$ is an integration constant (in the sense that its divergence is equal to zero).  It can be shown that this term is exactly zero for specific symmetries and that behaves at least as $r^{-3}$ for nonsymmetric configurations\cite{1984ApJ...286....7B}.  By assuming that the term is second-order, it is possible to obtain an algebraic equation for the spatial gradient of the modified potential, which is straightforward to solve once the Newtonian potential is calculated with the usual methods that are implemented in GR codes.  The method is exact for 1D or 2D codes respecting certain symmetries.  

 Note that while this method is approximate, it conserves the two main features of the MOND theories:  forces decaying as $1/r$ at large distances and a screening mechanism based on the gradient of the field.\footnote{In fact, the MOND community does not use the expression ``screening mechanism''.  Instead, they refer to the ``Newtonian limit'' of the theory.  However, these two terms make reference to the same thing:  the fact that these theories include mechanisms that can hide the MG effects when needed.  In the MOND case, the Newtonian limit (where the MG effects disappear) is obtained when the gradient of the MOND potential is large.}  The validity of the approximation for the highly nonsymmetric configurations associated with the cosmic web was tested in Ref.~\refcite{2008MNRAS.391.1778L} (unpublished results associated with this paper can be found also in 
\newpage
\noindent
Ref.~\refcite{2011PhDT.......253L}).  Among other things, it has been found that the curl term has the effect of boosting structure formation at intermediate scales.  Since this result was obtained in cosmological models that do not include dark matter (the dark matter effect is given by MOND), extrapolation of these results to the dark energy case must be made with care.  See Sec.~\ref{initial_guess} for a description of how this approximate solution can be used as an initial guess for iterative solvers of Eq.~(\ref{mond_equation}).

An alternative method is based on a different realization of the MOND theory (QU-MOND)\cite{2010MNRAS.403..886M}.  First simulations with this model were presented in Ref.~\refcite{2011PhDT.......253L}.  See also other implementations presented in Refs.~\refcite{2012MNRAS.421.2598A}, \refcite{2015MNRAS.446.1060C} and \refcite{2015CaJPh..93..232L}.  The gravitational potential of this theory is described by the following set of equations:
\begin{align}
\label{qumond_1}
\nabla^2\Phi_N & =  4\pi G \delta\rho \\
\label{qumond_2}
\nabla^2\Phi_M & =  \nabla\cdot\left[ \nu\left(\frac{|\nabla\Phi_N|}{a_0}\right)\nabla\Phi_N\right], 
\end{align}
where $\Phi_N$ is the Newtonian potential which is solution of the standard Poisson's equation, $\Phi_M$ is the MOND potential which will affect the trajectory of the particles and the $\nu$ free function is not to be confused with the $\mu$ interpolating function required in the AQUAL model.  Note that this representation of the MOND idea is straightforward to implement:  once the Newtonian potential was obtained by solving Eq.~(\ref{qumond_1}) with usual methods, it can be discretized on a grid and used as source of the Poisson's equation associated to the MOND potential (\ref{qumond_2}), which is also linear and can be solved with the same methods used for solving Eq.~(\ref{qumond_1}).

A different representation of the MOND idea can be derived from a bi-metric Lagrangian and was presented in Ref.~\refcite{2009PhRvD..80l3536M}.  This particular theory can be simulated by combining solutions of standard Poisson's solvers and nonlinear solvers of Eq.~(\ref{mond_equation}).  A particular implementation can be found in Ref.~\refcite{2011PhDT.......253L}.

\subsubsection{Code units}

The reference frame adopted in simulations is usually defined by the metric given by Eq.~(\ref{metric}) (i.e. comoving spatial coordinates with some particular choice of time).  On top of this, it is still necessary to choose units.  It is customary to describe the space and time coordinates in terms of the size of the box and the age of the universe ($1/H_0$).  The unit of mass is usually such that the mass of the particles is order one or that the mean density of the universe is exactly one.

In the MG case, it is also necessary to chose units for the extra degrees of freedom.  These are usually defined in terms of the background value of the field, its vacuum value or the Planck mass.  Rescaling the fields with a given power of the expansion factor may also help in removing some terms from the Klein--Gordon or geodesics equations.

Note that these units are \textit{not} natural units and thus, the speed of light and Planck constant are not equal to one in these simulations.  They must be included  \\\\
\noindent 
explicitly with the correct units or in the normalization of the fields.  This happens for instance when taking into account time derivatives in the equations for the extra degree of freedom.

\subsubsection{A word on performance}

Modified gravity simulations usually need to track additional degrees of freedom, so more calculations are likely to be required.  This will result in an overhead with respect to GR cosmological simulations, which in some cases may make MG simulations infeasible.  This overhead will naturally depend on the methods that are implemented, which in turn depends on the class of models we are dealing with.  The simplest models that can be simulated (i.e. those that only modify the background expansion or are condensed in a rescaling of the gravitational constant) have a straightforward implementation, which consists in substituting the tables that have the information about the background cosmology or the units associated to the force.  This kind of simulations has zero overhead and thus, can be run with the same resolution used for standard gravity simulations.  The same applies to nonlocal gravity (Sec.~\ref{section:non_local}) when appropriate approximations are made.

The situation is different for models which give rise to nonlinear elliptic equations (second- and third-classes of models presented in Table \ref{table:classification_methods}).  This kind of equations are typically solved with iterative methods.  The convergence rate strongly depends on the specifics of both the models and the solvers.  See for instance Ref.~\refcite{2017JCAP...02..050B} for an example of how different integration variables can have a dramatic change in the convergence rate of the solvers.  The overhead associated to these kind of simulations typically lie between a factor of 2 to 10 or 20.

Finally, the most complex models we can deal with give rise to hyperbolic equations (fourth-class of models in Table \ref{table:classification_methods}).  The only exact solvers that exist at present are conditionally stable, which means that they require a Courant--Friedrich--Levy condition on the time step.  The number of time steps associated to the MG solver increases by typically two to three orders of magnitude with respect to the standard gravity part of the code.  Implementation of implicit solvers for this family of models will largely increase the competitiveness of these simulations in terms of speed.

\section{Multigrid Methods for Nonlinear Partial Differential Equations}
\label{section:multigrid}

Most of the models summarized in the previous section are described by nonlinear hyperbolic equations (i.e. nonlinear wave equations).  This is different to what happens in the GR case, where an appropriate choice of the gauge can hide the time derivatives of the metric perturbations.  In order to keep the spirit of standard GR simulations (in which the metric perturbations are not evolved in time starting from some initial conditions, but calculated at every time step) it is usually assumed that the terms that include time derivatives in the MG equations are sub-dominant or important only in the background.  This is the so-called quasi-static \\ \\
\noindent
limit of the models, in which the additional degrees of freedom are described by elliptic equations.  The nonlinearity of these equations prevents us from applying techniques that rely on the superposition principle (such as Fourier-based solvers or direct summation of forces).  Instead, codes take advantage of multigrid techniques, which are the subject of this section.

To fix the notation, let us assume that we are interested in solving an equation that has the following form:
\begin{alignat}{2}
\label{eq_multigrid}
L(\partial^2\phi, \partial\phi, \phi) & = S(\rho, \phi) & \quad & \mathrm{in}~\Omega\\
\phi(x) & = f(x)                                        && \mathrm{in} ~ \partial\Omega
\end{alignat}
where $L$ is a differential operator (not necessarily linear), $S$ is a nonlinear scalar function of the density and the scalar field itself, $f$ is a function that defines the boundary conditions, $\Omega$ is the domain where the equation is defined and $\partial\Omega$ the boundary of $\Omega$.  We are interested in finding a numerical solution of this equation on a grid that covers the simulation domain.  The method consists in discretizing the equation on the grid and using an iterative method to obtain improved solutions starting from an initial guess.  In this section we will summarize different aspects of the iteration method as well as multi-level techniques for accelerating the convergence rate.

The multigrid techniques are described in great detail in Refs.~\refcite{Brandt77}, \refcite{Wesseling92}, \refcite{Trottenberg} and \refcite{brandt2011multigrid}.  Furthermore, a lot of information and tricks needed to solve MG equations can be found in the papers associated with specific implementations of the models summarized in Tables \ref{table:classification_methods} and \ref{table:classification_mond}.  Finally, as these methods are widely used in several other contexts, such as fluid dynamics, geophysics, medicine, etc., information may be found also in nonastrophysical journals.

\subsection{Gauss--Seidel iterations}
\label{subsection:gauss_seidel}

Before setting up an iteration scheme, it is necessary to discretize Eq.~(\ref{eq_multigrid}) on the grid.  Such discretization takes the following form:
\be
\label{eq_multigrid_discret}
L^l[\phi^l] = S^l[\rho^l, \phi^l].
\ee
The index $l$ represents the grid in which the problem is described.  The fields $\rho^l$ and $\phi^l$ are the density and the required solution on the grid.  Finally, $L^l$ is a discretization of the differential operator $L$.  Equation (\ref{eq_multigrid_discret}) represents a very large set of nonlinear algebraic equations, whose number is given by the number of nodes in the grid.  Our aim is to approximate the solution of the differential equation (\ref{eq_multigrid}) by the solution of the algebraic equation (\ref{eq_multigrid_discret}).

The discretization of the equation can be obtained by discretizing each derivative of the differential operator, writing them as linear combinations of values of the field at the neighbouring nodes.  For instance, second-order formulas for the second derivatives can be obtained using a 19-points stencil:
\begin{eqnarray}\label{eq:discexample}
\partial_x^2\varphi_{i,j,k} &=& \frac{1}{h^2} \left ( \varphi_{i+1,j,k} + \varphi_{i-1,j,k} - 2\varphi_{i,j,k} \right ) \\
\label{discretiz_xy}
\partial_x\partial_y\varphi_{i,j,k} &=& \frac{1}{4h^2} \left (\varphi_{i+1,j+1,k} - \varphi_{i+1,j-1,k} - \varphi_{i-1,j+1,k} + \varphi_{i-1,j-1,k} \right ),
\end{eqnarray}
where the indexes ${i,j,k}$ corresponds to the indexes of each grid in each Cartesian direction.

Equation (\ref{eq_multigrid_discret}) is solved iteratively with a Gauss--Seidel scheme.  In every iteration step of this scheme, the solution is updated in every element of the grid by finding the root of the following function
\be
\label{root_multigrid}
T^l[\phi^l] \equiv L^l[\varphi^l] - S^l.
\ee
In the linear case, it is possible to find an analytic expression for the updated value $\bar{\phi}^l$, which is a linear combination of the values of the field in the neighbouring nodes.  This gives rise to the so-called \textit{explicit} solvers.  For instance, the 1D standard Poisson's equation, can be discretized as follows:
\be
\frac{\phi_{i+1} - 2\phi_i + \phi_{i-1}}{h^2} = \rho_i.
\ee
In an explicit solver the solution in a given iteration step can be updated for instance in the following manner:
\be
\tilde{\phi}_i = \frac{1}{2}\left[ \phi_{i-1} + \phi_{i+1} - h^2\rho_i \right].
\ee
Exact details will depend on the type of discretization and exact algorithm used, but in general, each iteration step consists in updating the solution in every grid point $i$ with a linear combination of the values that surround $i$ and that correspond to the previous step.

In the nonlinear case, it is customary to use \textit{implicit} solvers, which consist in approximating the root of $T^l$ by applying one step of a Newton--Raphson algorithm:
\be
\bar{\phi}^l = \phi^l - \frac{T^l}{\frac{\partial T^l}{\partial \phi^l}}.  
\ee
Reference \refcite{2017JCAP...02..050B} has shown that implicit solvers are not the only route to deal with nonlinear problems.  It is in fact possible to obtain explicit algorithms by making appropriate changes of variables.

In order to fully specify a Gauss--Seidel scheme, we need to fix the order in which the iterations are made in each sweep through the grid.  For instance, in the lexicographic ordering, the update of the field in the cell $(i,j,k)$ is followed by a calculation in the cell $(i+1, j, k)$ and so on.  A better convergence rate and paralelization properties can be obtained by altering this scheme.  In the most popular scheme, each complete iteration step consists in two sweeps in which the update of a given cell $(i,j,k)$ is followed by an update on the cell $(i+2,j,k)$.  The two \\ \\ 
\noindent
sweeps start in different cells: $(0,0,0)$ and $(1,0,0)$ and thus, the scheme resembles a two colour chess board.  Similar schemes exists which consist in four or even eight colours.

\subsection{Multigrid acceleration}
\label{subsection:multigrid}

The quality of a given solution $\phi^l$ is usually determined by studying the amplitude of the residuals, which are defined as
\be
\epsilon^l =  L^l[\phi^l] - S^l[\rho^l, \phi^l].
\ee
The smaller the residual, the better the solution will be.  An associated quantity, which can be used to gauge the quality of the given solver, is the so-called convergence rate, which can be defined as the ratio of residuals at two consecutive iteration steps.  It can be shown through analysis of the convergence rate of individual Fourier modes \cite{Brandt77, Wesseling92, Trottenberg, brandt2011multigrid} that the Gauss--Seidel algorithm is very efficient in smoothing out the residuals, but has a very slow convergence rate when the solution includes modes whose wavelength is much larger than the spatial resolution of the grid.  This is due to the local nature of the method:  in each iteration step, information travels from each cell to its closest neighbours only and thus a very large number of iterations is required to transfer information across the domain.

It is possible to accelerate the convergence rate for these long modes by combining solutions obtained on coarser grids and coarser grids.\footnote{These different grid levels should not be confused with the different grid levels that exist in codes that include adaptive mesh refinements (AMR).  In that case, different grids are used to increase the resolution locally, while in the multigrid case described here, resolution is actually downgraded below the target resolution where we want to obtain the solution.}  The typical way of arranging these grids is to define a set of grids whose resolutions are one half, one quarter and so on of the target resolution. The coarsest grid of this set usually contains two nodes per dimension, but this is not a strict rule.  The coarsest grid can even be changed while the iterations progress.  In order to prove convergence properties of the method, it is enough to work with two grids.  The algorithm is the following:  a number of iterations is performed on the finest grid $l$ (i.e. the grid in which we want to obtain the solution).  When the convergence rate slows down (or a fixed number of iterations has been completed), iterations are made to approximate the error of this intermediate solution.  As most of the error will come from the large modes that did not converge yet, this is done on the next coarse grid $l-1$.  So the new equation for which iterations are made on the coarse grid is
\be
L^{l-1}[\delta\varphi^{l-1}] = R(\epsilon^l),
\label{eq_coarse}
\ee
where $R$ is a \textit{restriction operator} which transfers fields from the fine to the coarse grid.  Once a given number of iterations is made on the coarse grid, the additional 
\newpage
\noindent information that was collected related to the large modes is added to the solution on the fine grid:
\be
\bar{\phi}^l = \phi^l + P(\delta\phi^{l-1}),
\ee
The transfer between the coarse and fine grids is made with a \textit{prolongation operator} $P$.  See for instance Refs.~\refcite{Brandt77,Wesseling92,Trottenberg} or \refcite{brandt2011multigrid} for a detailed description of convergence properties of this method.

Application of the method to real problems requires repeating this process on more than one grid level.  By doing this, it is possible to reconstruct the solution on the finest grid by combining solutions of coarser and coarser grids which treat each set of Fourier modes separately.  The decision about changing the grid level can be made by taking into account the behaviour of the residuals and convergence rate or by using fixed schemes.  Popular schemes are the V cycle (in which the iterations are made starting from the finest grid, all the way down to the coarsest grid and then back to the finest grid) or the W cycle in which jumps between coarse levels are more often that between fine levels.  Well behaved solvers typically reduce the residual by one order of magnitude after each V cycle.

In the nonlinear case, it is not possible to add up solutions obtained in different grids and thus, the iterations in the coarse levels are not made to approximate the error $\delta\phi^{l-1}$, but the solution $\phi^{l-1}$ itself.  So Eq.~(\ref{eq_coarse}) is substituted by
\be
L^{l-1}[\varphi^{l-1}] = -R(\epsilon^l(\varphi^l, S^l)) + \epsilon^{l-1}(R(\varphi^l), R(S^l))
\ee
The correction then is made in the following way:
\be
\bar{\varphi}^l = \varphi^l + P(\varphi^{l-1} - R(\varphi^l)).
\ee
This extension to the method is known with the name of full approximation storage (FAS).

\subsection{Initial guess for the iterative solvers}
\label{initial_guess}

Making appropriate choices for the initial guess with which the iterations start can be crucial for reaching convergence in a reasonable number of iterations.  In general, this choice depends on the particulars of the equation to be solved.  The decision is usually made by trial and error.  The initial guess for obtaining solutions on the domain grid (i.e. the grid that covers the entire simulation box) can be obtained in the following ways:
\begin{itemize}[leftmargin=*]
\item Background value of the field.
\item Minimum of the effective potential.  
\item Extrapolation in time from previous time step.
\item Solution of auxiliary PDEs.  This method has been applied to the MOND equation (\ref{mond_equation}) in Ref.~\refcite{2011PhDT.......253L} (see Appendix A in this reference).  An auxiliary PDE can be constructed by taking the divergence of the field $\nabla\Phi_M$, which can be obtained after neglecting the curl term $\nabla\times \textbf{k}$ in Eq.~(\ref{spherical_mond}).  This is a linear Poisson's equation
\noindent
and thus, can be solved with Fourier-based algorithms which are both efficient and easy to implement.  When using this solution as the initial guess, the nonlinear multigrid solver that is required  to solve Eq.~(\ref{mond_equation}) does not need to make iterations to obtain the main features of the solution, but only needs to recover the effects associated with the curl field, which are small.
\end{itemize}
The initial guess for obtaining solutions in the refined levels is usually determined by interpolating values from the closest coarse grid.  This naturally applies only to AMR codes.

\subsection{Convergence criteria}

An important decision that needs to be made when developing a multigrid solver is related to the criterion that is used to stop the iterations.  Criteria that are too permissive may give inaccurate solutions, while criteria that are too strict may force the solver to make many more iterations than required to understand the underlying physics and may become extremely inefficient (especially in the context of $N$-body simulations, where the equations must be solved several hundreds or even thousands of times in a single simulation).  When designing a convergence criterion, there are three different solutions to take into consideration:
\begin{itemize}[leftmargin=*]
\item $\phi^l_n$ = approximate solution of the discretized equation (\ref{eq_multigrid_discret}) on the grid $l$ which was obtained after $n$ iterations.
\item $\phi^l$ = exact solution of the discretized equation (\ref{eq_multigrid_discret}) on the grid $l$.
\item $\phi$ = exact solution of the PDE (\ref{eq_multigrid}).
\end{itemize}
The following are important facts associated with these solutions:
\begin{itemize}[leftmargin=*]
\item The multigrid method converges after infinite number of iterations to the correct solution of the discretized equation:
\be
\phi^l_{\infty} = \phi^l.
\ee
\item The fact that the convergence rate $\eta$ in the grid $l$ is close to one (which means that the solution does not evolve when making iterations) does \textit{not} imply that the iterations have converged (it only implies that the iterations for the modes associated with the grid $l$ have converged):
\be
\left(\eta \equiv \frac{\phi^l_n}{\phi^l_{n+1}} \sim 1 \right) \nRightarrow \left( \frac{\phi^l_n}{\phi^l_{\infty}} \sim 1 \right).
\ee
\item The solution of the discretized equation is \textit{not} the solution of the differential equation evaluated at the nodes of the grid:
\be
\phi^l_{\infty} \neq \phi.
\ee
This is because the discretization of the equation introduces an error, which depends on the order of the discretization formulas used and on the resolution reached by the grid.  A commonly used estimation of the truncation error $\tau^l$ is\footnote{The truncation error can be estimated by comparing the result of applying the discretized differential operator in two consecutive grids $l$ and $l-1$ to the analytic solution of the equation:  $\tau^l = L^{l-1}\left[\phi\right] - L^{l}\left[\phi\right]$.  In practice, the analytic solution is substituted by the only solution that we have, which is the fine grid solution $\phi^l$.  In order to be able to apply the coarse grid differential operator $L^{l-1}$ to the fine solution $\phi^l$, we need to restrict it by applying the restriction operator $R$.  This gives rise to the first term of the right-hand side of Eq.~(\ref{def_tau}).  Similarly, once we applied the differential operator in both grids, we need to restrict the fine one to be able to make the comparison (which gives rise to the second-term of the right-hand side of this equation).}
\be
\tau^l = L^{l-1}\left[ R(\phi^l) \right] - R(L^l\left[\phi^l \right]).
\label{def_tau}
\ee 
\item The residual is not a dimensionless quantity, which means that it is possible to make it arbitrarily small by changing the units:
\be
[\epsilon^l] \neq 1
\ee
\item The residual is not necessarily uniform on the grid and may indeed have large variations across the domain:
\be
\epsilon^l(\phi^l_n) = \epsilon^l(\phi^l_n, x)
\ee
This is because the convergence rate is not uniform.  More complex features in the solution in specific regions will require more iterations.
\end{itemize}

Convergence criteria are usually based on amplitude of the residual, so the first thing we need to do is to fix is a norm for the residual.  Common choices are norm 2 or $\infty$:
\begin{align}
||.||^l_2 & = \Sigma_{i,j,k} \left( \epsilon_{i,j,k}^l \right)^2 \\
||.||^l_{\infty} & = \max |\epsilon_{i,j,k}^l|, 
\end{align}
where the indices $(i,j,k)$ run over all the grid points in the 3D box.

A few commonly adopted criteria follow:
\begin{itemize}[leftmargin=*]
\item Comparison with initial guess:
\be
\frac{||\epsilon_n||}{||\epsilon_0||} < a.
\ee
This criterion has the problem that if the initial solution is already good, then the solver will make more iterations than necessary.  It is possible to also have the opposite case, in which the initial solution is too bad.  In this case, reducing the residual by a given fraction of the initial residual, may not be good enough.  The technique is implemented in several solvers with standard or MG.  However it must be used with care.
\item Comparison with a fixed constant:
\be
||\epsilon_n|| < a.
\ee
This criterion has the problem that $a$ must be chosen according to the units used in the code.
\item Comparison with truncation error:
\be
\frac{||\epsilon_n||}{||\tau_n||} < a.
\ee
In this case, the iterations will stop once the error associated with the discretization of the equations dominates.
\end{itemize}
In all these criteria, $a$ is a small free parameter.

\subsection{Extended multigrid techniques}

Details of the solver strongly depend on the properties of the equation that we are dealing with.  Convergence properties depend on both the differential operator and the source of the equation.  In some cases it may be necessary to apply additional tricks to reach convergence in a reasonable number of iterations.  We describe a few examples.

\subsubsection{Selective iterations}

In situations in which the complexity of the solution strongly depends on the position in the domain (such as for instance systems in which the density is almost uniform in a large part of the domain and has strong variations in a small region) the solvers do not usually converge uniformly in the whole domain.  In some regions the solution will quickly approach the desired solution, while in others, it may require a much larger number of iterations to reduce the local residual.  In this cases, it is possible to implement \textit{selective iterations}, which means that some local criterion is established to stop the iterations.  Thus, the iterations will be made only in the region in which the solution has not converged and thus the overall computational effort will be reduced.

\subsubsection{Continuation method}

The continuation method can be applied to nonlinear equations for which a particular parameter defines a linear limit \cite{Brandt77}.  In these cases, it may be convenient to solve the linear equation in the first place and obtain intermediate solutions changing the values of the parameter such that the solution evolves from the linear case towards the fully nonlinear situation in a continuous way.

\subsubsection{Dealing with fully nonlinear equations:  Operator splitting}
\label{section:fully_non_linear}

Models that include Vainstein screening mechanism such as DGP or Galileon give rise to fully nonlinear partial differential equations for the additional degrees of freedom.  The equations are still second-order in space and time, but there is a nonlinear dependence with the second spatial derivatives, which represent a major challenge for multigrid solvers.  Naive multigrid implementations such as those described in Secs.~\ref{subsection:gauss_seidel} and \ref{subsection:multigrid} have poor convergence properties in high density regions, which make simulations impractical or simply impossible to run.  We describe here two different strategies that are commonly used to deal with this problem.  A comparison between different Vainstein solvers can be found in Ref.~\refcite{2015MNRAS.454.4208W}.

The first attempts to solve these equations implemented workarounds to this problem which consisted in using a spherically symmetric approximation (which gives equations that can be solved with a Fourier-based algorithm)\cite{2009PhRvD..80l3003S,2008PhRvD..77b4048L} or by smoothing out the density distribution before sending it to the multigrid solver\cite{2009PhRvD..80d3001S,2009PhRvD..80l3003S,2009PhRvD..80f4023K}.  The second approach can improve the convergence rate and make the fully nonlinear simulations feasible.  However, it comes with a price in accuracy.

An alternative method, the operator splitting technique, was proposed in Ref.~\refcite{2009PhRvD..80j4005C} and implemented afterwards in the code ECOSMOG\cite{2013JCAP...05..023L, 2013JCAP...11..012L}.  Let us fix notation by assuming that we are interested in finding solutions of the following equation:
\be
\nabla^2\phi - A\left[ \left( \nabla^2\phi \right)^2 - (\partial_i\partial_j\phi)(\partial^i\partial^j\phi) \right] = B \delta\rho, 
\label{eq_dgp_for_splitting}
\ee
which has the same form of the DGP equation of motion and where $A$ and $B$ are model dependent constants and $\partial_i$ are partial derivatives with respect to spatial coordinates.  In order to define the Gauss--Seidel iterations, we will need to solve a discretized version of the equation for the field $\phi_{l,m,n}$ in each node $(l,m,n)$.  If we can ensure that the discretization of the term $(\partial_i\partial_j\phi)(\partial^i\partial^j\phi)$ in the node $(l,m,n)$ is independent of $\phi_{l,m,n}$, then we can see the equation as an equation for $(\nabla^2\phi)_{l,m,n}$ and solve it analytically for $\phi_{l,m,n}$ once we obtained $(\nabla^2\phi)_{l,m,n}$.  This can be done by decomposing the term $\partial_i\partial_j\phi$ into trace and traceless parts:
\be
\partial_i\partial_j\phi = T_{ij} + S_{ij}, 
\ee
where
\begin{align}
T_{ij} & = \frac{\nabla^2\phi}{3} I \\
S_{ij} & = \partial_i \partial_j \phi - \frac{\nabla^2\phi}{3} I, 
\end{align}
where $I$ is the identity operator.  Substitution of these definitions in $(\partial_i\partial_j\phi)(\partial^i\partial^j\phi)$ gives
\be
(\partial_i\partial_j\phi)(\partial^i\partial^j\phi) = \left( \frac{\nabla^2\phi}{3}\right)^2 + \left[ 2S_{ij} T^{ij} + S_{ij}S^{ij} \right],
\ee
which in turn gives the following alternative version of the equation of motion (\ref{eq_dgp_for_splitting}):
\be
\nabla^2\phi - A\left\{ \frac{2}{3}\left( \nabla^2\phi \right)^2 - \left[ 2S_{ij} T^{ij} + S_{ij}S^{ij} \right] \right\} = B \delta\rho.
\label{dgp_nice}
\ee
The term inside the square brackets in this expression fulfils the property that we are interested in:  the usual discretization that can be obtained in the cell $(l,m,n)$ by applying Eq.~(\ref{discretiz_xy}) is independent of the value of the solution $\phi_{l,m,n}$ in that cell.

It has been found that this way of arranging the terms largely improves the convergence rate of multigrid solvers.  However, it does not solve all the issues with this model since the solution of Eq.~(\ref{dgp_nice}) for $\nabla^2\phi$ may give rise to square roots of negative numbers in under-dense regions, where density perturbations are negative.  This does not occur in the DGP case, but 
in the quartic Galileon\cite{2013JCAP...11..012L,2013JCAP...11..056B} model.  See Ref.~\refcite{2015PhRvD..92f4005W} for a detailed discussion on this issue.

\subsection{Summary}

This section summarizes the decisions that need to be taken when developing multigrid codes.  This list of decisions should not be considered as a simple summary, but as a checklist of properties of the solver that can be revised when things do not go as expected.  So if the solver that the reader is developing has poor convergence rate, it may be worth checking the following properties of his or her implementation:
\begin{itemize}[leftmargin=*]
\item Parametrization of the theory:  Can we find an alternative representation of the theory which simplifies the equations?
\item Integration variables (independent variables as well as fields).
\item Stencil and discretization formula for the PDE inside the domain.
\item Stencil and discretization formula for the PDE at the boundaries.
\item Initial guess for the iterations.
\item Type of iterations (explicit, implicit, over-relaxation).
\item Sweeping strategy (2, 4, 8 colors?).
\item Number of multigrid levels.
\item Number of iterations per multigrid level (which may be different from level to level).
\item Multigrid scheme ($V$, $W$, adaptive, etc.).
\item Restriction and prolongation operators.
\item FAS versus no FAS scheme.
\item Convergence criterion.
\item Extended techniques:  selective iterations, continuation methods, operator splitting, etc.
\end{itemize}
Taking these decisions constitutes a form of art, whose outcome depends on the particular details of the equations to be solved.  What works for one equation may not work for others.

Once a newly implemented solver can pass several tests and is accepted as being accurate enough, it may be a good idea to optimize it.  A few ways in which this can be done are discussed in Refs.~\refcite{2015JCAP...12..059B} and \refcite{2017JCAP...02..050B}.  The first of these optimization methods consists in reducing the spatial resolution when solving the Klein--Gordon equation for the additional degree of freedom.  The impact of resolution in these solutions was also discussed in Ref.~\refcite{2014PhRvD..89h4023L}.

In cases in which the code needs to solve both, the Poisson's equation for the metric perturbation and the Klein--Gordon equation for the scalar field, it is possible to increase the speed of the simulations by solving both equations in parallel.  As the GR and MG solvers work with very similar routines, it is possible to share the loops that have to be made through the grid to calculate different quantities.  This method was implemented in the code Isis \cite{2014A&A...562A..78L} for the symmetron model and it was found to give an improvement in speed of about 20\%.

\section{Nonstatic Solvers for Nonlinear Partial Differential Equations}
\label{section:non_static}

The dynamics of MG degrees of freedom is usually described by nonlinear hyperbolic equations (please, see Refs.~\refcite{weinberger1995first} or \refcite{agarwal_equations}, your favourite book on partial differential equations or the notes of your undergrad courses if in need of refreshing classification of equations).  A common practice is to assume the so-called quasi-static approximation, which consists of assuming that the time derivatives of these fields are small or important only in the background.  This approximation usually gives rise to nonlinear elliptic equations, which can be solved with the methods described in Sec.~\ref{section:multigrid}.  This section is about going beyond this approximation.

\subsection{The quasi-static approximation}

Terms that include time derivatives of the fields are not exclusive to the field equations associated with MG.  For instance, in the GR case, it is necessary to deal with time derivatives when deriving the usual Poisson's equation from basic principles.  The effects associated with these terms were analysed in Refs.~\refcite{2011PhRvD..83l3505C},\refcite{2012PhRvD..85f3512G},\refcite{2015PhRvD..92l3517F} and \refcite{2017JCAP...12..022F}.

The validity of the quasi-static approximation in a MG context was discussed for first time in Ref.~\refcite{2008PhRvD..78l3523O}.  In that early work, the time derivatives were not take into account when running the simulations, but estimated afterwards by post-processing the simulation data (i.e. by applying appropriate discretization formulas which combine static solutions that were obtained at different time steps).  The results show that the terms that include time derivatives in the $f(R)$ equations are sub-dominant when compared with the spatial derivatives.  A similar comparison was made by myself using symmetron simulations that were presented in Ref.~\refcite{2013PhRvL.110p1101L}.  I found that, while it is true that globally the time derivatives are sub-dominant, they may be larger than the spatial derivatives when restricting the analysis to under-dense regions (i.e. voids), where the Laplacian of the field is very small.

These results can only be confirmed by running simulations beyond the static limit and comparing the data with the output of static codes, which do not rely of the quasi-static approximation.  This has been done in the linear regime for the $f(R)$ model\cite{2014PhRvD..89b3521N} and in the nonlinear regime for two different models:  symmetron and $f(R)$.\cite{2013PhRvL.110p1101L,2014PhRvD..89h4023L,2015JCAP...02..034B,2015MNRAS.454.4208W}.  The approximation was found to be good enough for predicting the matter and velocity power spectrum.  For instance, in the case of the symmetron model, a difference of 0.2\% was found between static and non-static power spectra in a range of scales from $k$=0.05 to $k$=10 $h$/Mpc\cite{2014PhRvD..89h4023L}.  Differences in the mass function were studied in Ref.~\refcite{2015MNRAS.454.4208W} and also found to be well below 1\% for masses between $3\times10^{12}$ and $10^{14}$.  In the case of $f(R)$ model, differences between static and nonstatic results are within the same ranges\cite{2015JCAP...02..034B}.  However, it is worth keeping in mind that these nonlinear codes contain simplifications.  Furthermore, the impact of these additional terms on different observables is still unknown.

It is worth noticing that there are models, such as disformal gravity, in which the time derivatives of the field appear not only in the kinetic part of the equations of motion, but also in the coupling to matter.\cite{2016A&A...585A..37H}.  
This provides a very rich phenomenology which does not exist in models that are conformally coupled.  Neglecting time derivatives when running simulations with this kind of model may prevent us from predicting the unique signature of new physics that we are looking for.  So nonstatic simulations are crucial for testing these models.

A problem that arises when running nonstatic simulations is related to the determination of initial conditions for the additional degrees of freedom.  Models that include screening mechanisms are expected to give rise to small perturbations at the start of the simulations.  However, the exact shape and amplitude of the spectrum of these perturbations may still be important.  This issue was discussed in Ref.~\refcite{2016A&A...585A..37H}, where it was shown that the end result of the simulations is independent to the amplitude that is chosen for the initial perturbations of the field.  In this particular model (disformally coupled symmetron), this is not only a consequence of the screening, but also of the fact that there is a phase transition at a redshift of about one.  The scalar field receives a large kick when this happens, which forces it to loose memory of any small perturbations that may have existed at early times.  Note that this result may not be valid for models that do not include screening of the field at high redshift.  In this case, initial conditions will have to be generated more carefully, for instance, as a realization of a power spectrum generated with linear theory.

\subsection{Solving hyperbolic equations in $N$-body simulations}

The first solvers for hyperbolic equations in the context of MG $N$-body simulations were applied to the symmetron model and are based on the leap-frog algorithm\cite{2013PhRvL.110p1101L,2014PhRvD..89h4023L}.  The method consists in converting the second order Klein--Gordon equation into a first order system of equations with the following change:
\be
q = a^b \dot{\phi}, 
\ee
where $b$ is chosen such that commonly existing terms cancel.  The resulting equation has the following form:
\begin{align}
\dot{\phi} & = \frac{q}{a^b}\\
\dot{q} & = S(\rho, \phi),
\end{align}
where $S$ is a model dependent function.  Note that these equations are schematic and that details may depend on the model.  These equations are similar to Hamilton equations for a set of particles with the difference that instead of evolving the position of the particles, they evolve the amplitude of the field at every point of space.  So by discretizing the field on a grid, it is possible to apply the same algorithm that is commonly used to integrate Hamilton's equations in $N$-body codes.  The method was applied to uncoupled scalar fields within GR for instance in Refs.~\refcite{1989PhRvD..40.1002W} and \refcite{1989ApJ...347..590P}.  Implementation to coupled scalar fields can be found in Refs.~\refcite{2013PhRvL.110p1101L}, \refcite{2014PhRvD..89h4023L} and \refcite{2016A&A...585A..37H}.  As in the model studied in these papers the scalar field undergoes very fast oscillations, the simulation has to deal with two different time scales (one associated with normal gravity and a much shorter one associated with the scalar field).  This problem has been solved by having two different time steps in the simulation: a main time step as in a usual GR code and a much smaller one for modified gravity.  As the time step that is employed to move the particles is the one associated with GR, these simulations still have the limitation that the particles can not see the oscillations of the scalar field and so the impact of the scalar waves in the matter distribution is still an open question.  The same method has been implemented in 1D in spherical coordinates\cite{2015PhRvD..92f4005W,2017PhRvL.118j1301H} or 2D cartesian grid\cite{2014PhRvD..90l4041L}.

The leap-frog algorithm has the disadvantage that it is conditionally stable, which means that it requires a Courant--Friedrich--Levy condition on the time step:  For the method to work, the time step has to be such that no information is propagated in more than one spatial cell element within the step\cite{TORO:60926, Press:2007:NRE:1403886}.  While this condition does not affect the quality of the results, it forces the simulation to work with extremely small time steps, which makes them very expensive.  A work around to this problem consists of introducing numerical viscosity which suppress small scale oscillations of the solution and can be done by means of implicit methods.

A different implementation of nonstatic solvers was presented in Ref.~\refcite{2015JCAP...02..034B} and applied to the $f(R)$ model.  In this case, the time derivatives of the field were estimated with a down-wind discretization:
\begin{align}
\left(\frac{\partial\phi}{\partial t}\right)_n & = \frac{\phi_n - \phi_{n-1}}{t_n-t_{n-1}}, \\
\left(\frac{\partial^2\phi}{\partial t^2}\right)_n & = \frac{({d \phi}/{dt})_n - (d \phi/dt)_{n-1}}{t_n-t_{n-1}}, 
\end{align}
which involves information from a given time step $n$ and the previous one $n-1$.  Once this is done, it is possible to translate the terms that depend on these derivatives to the right-hand side of the Klein--Gordon equation:
\be
\nabla^2\phi_{n+1} = f(\phi_{n+1}, \rho_{n+1}) + g(\dot{\phi}_n, \ddot{\phi}_n), 
\ee
where $f$ and $g$ are model dependent functions.  This way of discretizing the Klein--Gordon equation gives rise to an elliptic equation for the scalar field at the time step $n+1$ and, which can be solved with static solvers described in Sec.~\ref{section:multigrid}.  Detailed discussion of up-wind and down-wind discretizations can be found for instance in Refs.~\refcite{TORO:60926} and \refcite{Press:2007:NRE:1403886}.

\newpage
Additional information about integration of nonstatic equations in gravity can be found in the GR literature.  Refs.~\refcite{2016JCAP...07..053A}, \refcite{2017JCAP...11..004A} and \refcite{2017arXiv170904703L} present examples of implicit and explicit solvers that can be used for going beyond the static limit in GR.  Furthermore, as discussed in Sec.~\ref{section:multigrid}, the equations associated with MG are not unique to gravity, but are used also in different contexts such as geophysics, engineering (especially in hydrodynamical applications) or medicine.  Communication with these communities is essential for making progress in this field.

\section{Baryonic Simulations}

The only difference that exists between simulating dark matter and simulating baryons is that in the baryonic case, it is not possible to neglect nongravitational interactions between the particles.  This gives rise to a plethora of thermodinamic and quantum processes between different elements of the periodic table, transitions between them, transfer of energy between these elements and photons, etc.  The complexity of the solutions when including all these effects increases disproportionally:  A simulation that includes these interactions and which has infinite resolution may end up producing humans that may run simulations which may generate other humans that may run other simulations and so on.  Naturally, we do not have the computational resources that are required to run such a simulation, so we need to make simplifications.  These simplifications are based on the idea of dividing the solutions of the equations into different sets of scales.  For the large scales, the equations are solved self-consistently to follow the evolution of dark matter and baryonic perturbations.  The effects that occur at scales that are smaller than the resolution of the simulation are treated analytically following different processes with approximate models.  As we are not interested in studying the formation of humans in a cosmological context but only galaxies, these sub-grid recipes have also a cut off in resolution.

The complexity of baryonic simulations increases as the field progresses.  The first baryonic simulations included only adiabatic gas and no sub-grid physics.  Later on, cooling of the gas was included, followed by recipes for stellar formation, stellar evolution and different kinds of feedback (from supernovae, black holes, etc).  The state-of-the-art today includes magnetic fields.  Information about these kind of simulations within GR can be found in the presentation of some of the latest simulations and references there in\cite{2013MNRAS.436.3031V, 2014MNRAS.444.1518V, 2015MNRAS.446..521S, 2018MNRAS.473.4077P}.  The following sections describe how to include MG in these simulations.

\subsection{Including MG in hydrodynamical solvers}

In order to understand the evolution of baryons in the nonlinear regime, we need to reformulate the definition of cosmological simulation provided in Sec.~\ref{section_nbody_eqs} (Definition~\ref{definition_1}).  In particular, we need to relax the assumption of collisionless matter that is encoded in the energy-momentum tensor.  Doing this will prevent us from describing the fluid by following geodesics (Eq.~(\ref{eq_geo})).  So to derive alternatives to these equations, we need to repeat the analysis that was presented in Secs.~\ref{section_gr} and \ref{section_mg}, which resulted in Definition \ref{definition_2}.  I provide a schematic of how this can be done below.  More details on these derivations can be found in the works that describe specific implementations\cite{2011MNRAS.412..911C, 2013MNRAS.436..348P, 2015MNRAS.449.3635H, 2016PhRvD..93l3512H}.

Let us assume, for instance, a gravitational model described by the Lagrangian (\ref{general_lagrangian}).  Minimizing the action with respect to the metric will give rise to the following equation:
\be
G_{ab} = \frac{\delta \tilde{g}_{ab}}{\delta g_{ab}} \tilde{T}_{ab} + T^{\phi}_{ab}, 
\label{einstein_general}
\ee
where the two energy--momentum tensors are the usual Jordan frame matter tensor and the Einstein frame energy--momentum tensor of the scalar field:
\begin{align}
\label{energy_momentum}
\tilde{T}_{ab} &\propto \frac{1}{\sqrt{-g}}\frac{\delta S_M}{\delta \tilde{g}_{ab}}, \\
T^{\phi}_{ab} &\propto \frac{1}{\sqrt{-g}}\frac{\delta S_{KG}}{\delta g_{ab}}.  
\end{align}
Taking the divergence of Eq.~(\ref{einstein_general}) gives a generalized equation of conservation of energy.  After projecting the result in the time and space directions, we will end up with the usual continuity and Euler equations that can be found in hydrodynamical codes plus MG terms that involve derivatives of $T^{\phi}_{ab}$.  In the Einstein frame, these new terms take the form of an additional energy in the continuity equation and a fifth force in the Euler equation and depend on the scalar field itself and its derivatives in space and time.  Thus, they can be obtained by solving the Klein--Gordon equation for the field with the methods described in Sec.~\ref{section:multigrid} or \ref{section:non_static}.

According to this analysis, implementing MG hydrodynamical solvers may be straightforward:  a MG hydrodynamical solver can be obtained simply by plugging the solution of the Klein--Gordon equation that already exists in collisionless MG codes into a standard hydrodynamical solver.  However, additional complexity may arise, which could lead to the need for more complex implementations.  For instance, in the case where the free function $f$ in Eq.~(\ref{general_lagrangian}) corresponds to a standard kinetic term, the energy--momentum associated to the scalar field is
\be
T^{\phi}_{ab} = \nabla_a\phi \nabla_b\phi - \frac{1}{2}g^{cd}\nabla_c\phi\nabla_d\phi g_{ab} - V(\phi)g_{ab}, 
\ee
which is diagonal for a diagonal metric (i.e. when neglecting vector and tensor modes).  However, in the more general case in which $f$ includes cross derivatives, this assumption may be broken, which may force us to take into account nondiagonal components of the equations.  Also the Klein--Gordon equation for these models is sourced by the trace of the energy--momentum tensor.  In the case of baryons, there will be a pressure term sourcing the equation, which is not included in existing simulations.

\newpage
\subsection{Including MG in sub-grid models and semi-analytics}

The sub-grid models contain information about baryonic processes that occur at the very small scales that lie below the resolution of the self-consistent part of the simulation.  These processes are certainly affected by the presence of gravity and thus, depend on the assumed gravitational theory.  For instance, the formation of stars depends on how gas clouds fragment and collapse under their own gravity.  Different gravitational theories will give rise to different evolution of the clouds.  This in turn should give rise to a different initial mass function, which is one of the ingredients in these sub-grid models.  In present simulations, these extremely complex gravitational effects are included implicitly in the models (i.e. in most of the cases, the gravitational constant does not appear explicitly in the equations).  These highly nonlinear phenomena are described by simple equations and condense their complexity into a set of few free parameters.  In the GR case, these parameters are chosen by trial and error:  Large number of test runs is performed varying model parameters until a few observables such as the stellar formation rate density, the luminosity function of galaxies, etc. agree with the observed quantities.  This process of tuning the parameters of the simulations is very costly and done by large collaborations.

The generalization of the sub-grid models to MG is a two step process.  First it is necessary to re-derive the analytic expressions that describe different aspects of the models.  Second, the model parameters have to be recalculated.  In practice, constructing these models and tuning the simulations is an extremely complex task, which can not be done by the small MG community.  The state-of-the-art simulations with MG do not include specifics of MG in the definition of the models and instead use sub-grid models and parameters given by the GR simulations that are usually run for comparison.  Thus, existing simulations are not yet self-consistent.

For models that include screening mechanisms, the approximation that is made when not redefining the sub-grid models might be justified by the fact that the impact of sub-grid physics is dominant in high density regions, where the additional degrees of freedom should be screened.  Examples of these kind of simulations can be found in Refs.~\refcite{2014MNRAS.442..176P} and \refcite{2016PhRvD..93l3512H}.  The impact of MG in some of the small scale problems that the $\Lambda$CDM model historically faces, and whose proposed solutions within GR are usually related to the impact of baryons, was presented in Refs.~\refcite{corbett_moran_simulations}, \refcite{2015MNRAS.453.1371M} and \refcite{2016MNRAS.461.2490P}.

A different approach usually applied in a GR context to the determination of statistical properties of galaxies consists in taking into account baryonic effects only in the post-processing phase of the simulations.  In this case, the simulations follow only the dark matter component.  A set of recipes is applied \textit{a posteriori} on the resulting merger trees of dark matter halos to determine the properties of the galaxies that form inside them.  A change in the underlying gravitational model, will affect the outcome of these calculations through differences in both the merger trees and details on the semi-analytic model itself.  As in the case of sub-grid physics used in full hydro simulations, applying GR semi-analytic codes to MG merger trees may not give realistic results, specially in the MG models that include coupling.  Several quantities have to be redefined.  A simple example is given by the hydrostatic equilibrium, which is obtained by opposing pressure forces to gravity forces and which is usually part of the semi-analytic models.  In the MG case, the fifth force should be included in the analysis.  Another example is the distribution of spin parameter of the dark matter halos.  We know from simulations that the distribution in MG is log-normal as in the GR case, but the parameters (which are not free, but obtained from simulations) are different.  So a correct implementation of a semi-analytic model in MG will require changing these and other similar quantities in the codes.  Furthermore, retuning of the free parameters of the models will be required.  See Refs.~\refcite{2013MNRAS.436.2672F} and \refcite{2015MNRAS.452..978F} for examples of MG gravity implementations.

The question of whether the differences found between GR and MG galaxy formation simulations are real or a consequence of poorly tuned baryonic models is still open.  For instance,  differences were found in rotation curves of galaxies in Ref.~\refcite{2014MNRAS.442..176P}.  Are these differences based on reality or simply on the fact that the simulations need re-tuning?  Can we recover the GR rotation curves in a MG context by retuning parameters?  Only by re-running the simulations we will be able to give answers to these questions.  This is extremely challenging, but by doing it we may open the door for establishing new tests which may give us the definitive answer about gravity.  The risk is high, but certainly worth taking.

\section*{Acknowledgements}

CLL acknowledges support from STFC consolidated grant ST/L00075X/1 \& ST/P000541/1 and ERC grant ERC-StG-716532-PUNCA.  CLL thanks Shaun Cole for carefully reading the manuscript and anonymous referees and editor whose comments greatly improve the quality of the paper.

\bibliographystyle{ws-ijmpd}
\bibliography{references_claudio}

\end{document}